\documentclass[]{spie}  

\usepackage{amsmath,amsfonts,amssymb}
\usepackage{graphicx}
\usepackage[colorlinks=true, allcolors=blue]{hyperref}
\usepackage{float}

\title{Superpolished OAPs for WFIRST CGI}

\author[a]{M\'elanie Roulet}
\author[a]{Emmanuel Hugot}
\author[b]{Carolyn Atkins}
\author[a]{Sabri Lemared}
\author[a]{Simona Lombardo}
\author[a]{Marc Ferrari}
\affil[a]{Aix Marseille Univ, CNRS, CNES, LAM, Marseille, France}
\affil[b]{UK Astronomy Technology Centre, Royal Observatory, Edinburgh, EH9 3HJ, UK}

\authorinfo{melanie.roulet@lam.fr}

\begin{document} 
\maketitle

\begin{abstract}
Exoplanet imaging requires super polished off-axis parabolas (OAP) with the utmost surface quality. In this paper we describe an innovative manufacturing process combining 3D printing and stress polishing, to create a warping harness capable of producing any off axis parabola profile with a single actuator. The warping harness is manufactured by 3D printing. This method will be applied to the production of the WFIRST coronagraph's off axis parabolas. The evolution of the warping harness design is presented, starting from a ring warping harness generating astigmatism, to an innovative thickness distribution harness optimised to generate an off axis parabola shape. Several design options are available for the prototyping phase, with their advantages and disadvantages which will be discussed.

\end{abstract}

\keywords{Off axis parabolas, stress polishing, 3D printing, warping harness, finite element analysis}

\section{INTRODUCTION}
\label{sec:intro}  

The Wide Field Infrared Survey Telescope (WFIRST) is the next NASA flagship mission beyond the James Webb Space Telescope (JWST) \cite{2015arXiv150303757S}. The telescope is scheduled to launch in 2024 to observe exoplanets and high red-shift galaxies formed just after the Big Bang. In this paper we are interested in the coronagraph instrument (CGI) which is capable to perform direct imaging for exoplanet detection. The principle is to occult the star with a mask to create a disk where exoplanets are detectable. Many optical components are used within the  chronograph to achieved high contrast imagery and in this paper we are interested in the eight off axis parabolas.

Off axis parabolas are used in this coronagraph as a relay. They are located between the other optical components to relay the beam between the focal plane to the pupil plane. The surface of each off axis parabola is required to be superpolished to minimise the introduction of wave-front errors in the beam transmission. 

In this paper we describe the development of a new manufacturing process for off axis parabolas. Using stress polishing we can reach very high quality surface, perfectly suited for high contrast imaging. Using 3D printing we can create an innovative warping harness which is able to generate an off axis parabola shape. These two techniques are combined to develop an easy manufacturing process for off axis parabolas. 
  
\section{Off axis parabolas definition}
The Zernike decomposition is used to define the circular surface shape of the off axis parabolas\cite{Noll:76}. The first seventeen Zernike aberrations are listed in the Table \ref{coeff}, they are sorted with $n$, the radial coefficient increasing and $m$, the azimutal coefficient increasing. 
 
\begin{table}[H]
\begin{center}
\begin{tabular}{|c|c|c|c|c|}
  \hline
  Number & Aberration & Expression & n & m\\
  \hline
  1 & Piston & 1 & 0 & 0\\
  2 & Tilt x & $\rho \cos\theta$ & 1 & 1\\
  3 & Tilt x & $\rho \sin\theta$ & 1 & -1\\
  4 & Spherical & $2\rho^2 -1$ & 2 & 0\\
  5 & Astigmatism 3$x$ & $ \rho^2 \cos (2\theta)$ & 2 & 2\\
  6 & Astigmatism 3$y$ & $ \rho^2 \sin (2\theta)$ & 2 & -2\\
  7 & Coma 3$x$ & $(3\rho^2 -2)\rho \cos\theta$ & 3 & 1\\
  8 & Coma 3$y$ & $(3\rho^2 -2)\rho \sin\theta$ & 3 & -1\\
  9 & Trefoil 5$x$ & $ \rho^3 \cos (3\theta)$ & 3 & 3\\
  10 & Trefoil 5$y$ & $ \rho^3 \sin (3\theta)$ & 3 & -3\\
  11 & Spherical 3 & $6\rho^4 - 6\rho^2 + 1$ & 4 & 0\\
  12 & Astigmatism 5$x$ & $(4\rho^2 -3)\rho^2 \cos(2\theta)$ & 4 & 2\\
  13 & Astigmatism 5$y$ & $(4\rho^2 -3)\rho^2 \sin(2\theta)$ & 4 & -2\\
  14 & Squad 7$x$ & $\rho^4 \cos (4\theta)$ & 4 & 4\\
  15 & Squad 7$y$ & $\rho^4 \sin (4\theta)$ & 4 & -4\\
  16 & Coma 5$x$ & $(10\rho^4 -12\rho^2 +3)\rho \cos\theta$ & 5 & 1\\
  17 & Coma 5$y$ & $(10\rho^4 -12\rho^2 +3)\rho \sin\theta$& 5 & -1\\
  \hline
\end{tabular}
\caption{Zernike Aberrations}
  \label{coeff}
  \end{center}
\end{table}

\begin{figure}[H]
  \centering
  \includegraphics[width = 0.7\linewidth]{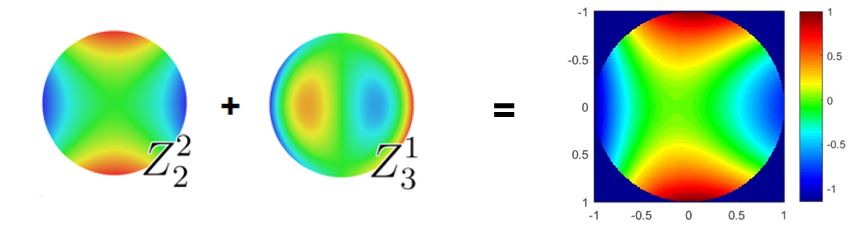}
  \caption{Off axis parabola definition using Zernike aberrations, with a deformation of 15\% Coma}
  \label{def}
\end{figure}

An off axis parabola can be defined by Zernike aberrations as a combination of astigmatism 3 and coma 3. Figure \ref{def} presents the shape of an off axis parabola with a $Ratio_{coma/astig}$ of 15\% using Zernikes. This definition will be used in all this paper. 

\section{WFIRST Requirements}

\begin{figure}[H]
  \centering
  \includegraphics[width = 0.7\linewidth]{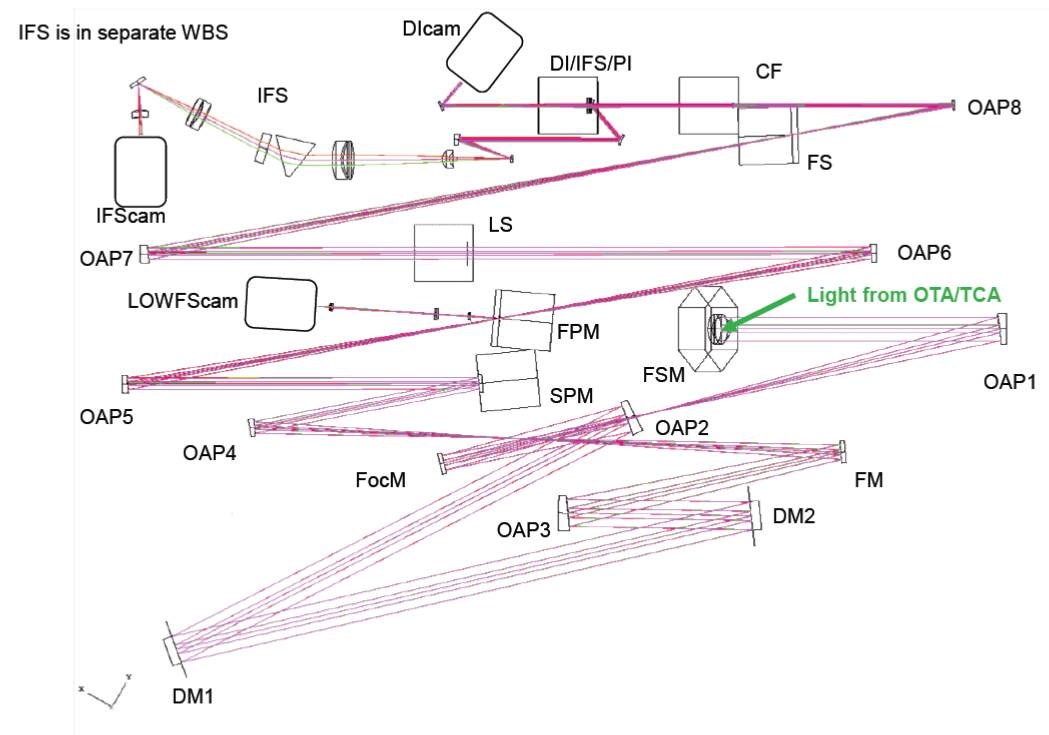}
  \caption{Optical design of the WFIRST coronagraph. Courtesy: Zhao Feng, NASA/JPL. OAP: Off Axis Parabola, DM: Deformable mirror}
  \label{corona}
\end{figure}

The coronagraph instrument is composed by eight off axis parabolas as shown in Figure \ref{corona}. Each parabola has different optical prescription in terms of astigmatism and coma. The average $Ratio_{coma/astig}$ is approximately 10\% for all eight optics, this ratio will be our requirement in this paper for the following study. 

\section{Off axis parabolas new manufacturing process}

\subsection{Stress polishing technique}
The idea is to use stress polishing to imprint the off axis parabola shape on the top surface of the substrate. Stress polishing is a technique developed by the German astronomer Bernhard Schmidt in the 1930s \cite{Lemaitre:72}. The process consists in applying forces through a warping harness to generate the off axis parabola shape on the substrate while polishing the part as a spherical surface. During this step the deformation created by the warping harness is imprinted on the substrate surface. After the polishing phase, the warping harness is removed and the substrate comes back to its initial position. To use this process the warping harness must generate a warping function equal to the inverse of the required final shape (i.e the off axis parabola) and the forces applied must be under the yield strength limit.

\begin{figure}[H]
  \centering
  \includegraphics[width = 0.5\linewidth]{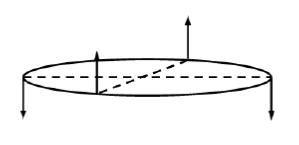}\hfill
  \includegraphics[width = 0.4\linewidth]{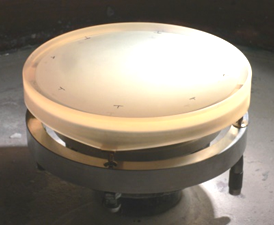}
  \caption{Left: Picture of the combination of forces applied to obtain pure astigmatism deformation.Right: Picture of the substrate, warping harness and the ring applying force}
  \label{stress}
\end{figure}

This technique provides several advantages. We can manufacture low cost - high performance optics, as long as the mechanical deformation is handled, precisely controlled and within the breakage limit of the mirror substrates.  Stress polishing is suitable to provide high quality surface because the dimensions of the polishing tool and the mirror are equal, which avoids high frequency in the surface error \cite{E2018}.

\subsection{3D printing technique}

In this study the warping harness will be manufactured with 3D printing. This fabrication process allows an innovative design with hollow structures and non-standard thickness distribution. The design constraints are different from traditional manufacturing and this gives us the opportunity to explore a new range of structures. 
Initially developed with plastics, 3D printing now offers a wide range of materials from ceramics to metals as well as composite materials. As discussed in Section \ref{sec:proto} promising ceramics materials, such as Cordierite, will be investigated.

The new fabrication process for off axis parabolas, 3D printing combined with stress polishing, does not impact the final error budget of the optical surface.
The error due to the warping harness printing is under $100\mu m$ Root Mean Square (RMS). This precision depends of the laser beam diameter and the layer thickness. We create a force application system composed by one actuators and micrometric screws with high transmission factor (around 1000 times). Combining 3D printing and this force application system with spherical polishing provides the high quality surface desired by the WFIRST optical requirements.

\section{HARNESS DESIGNS STUDY}

In this section the evolution of the warping harness will be presented, starting from an astigmatism mirror and developing to an non symmetric design with a thickness distribution optimised for the required off axis parabola profile.

\subsection{Astigmatism mirror}
A simple application of stress polishing is the astigmatism mirror. We applied two opposite forces as shown in Figure \ref{stress} (left). The warping harness is a simple ring as shown in Figure \ref{stress} (right). 

\begin{figure}[H]
  \centering
  \includegraphics[height =5cm]{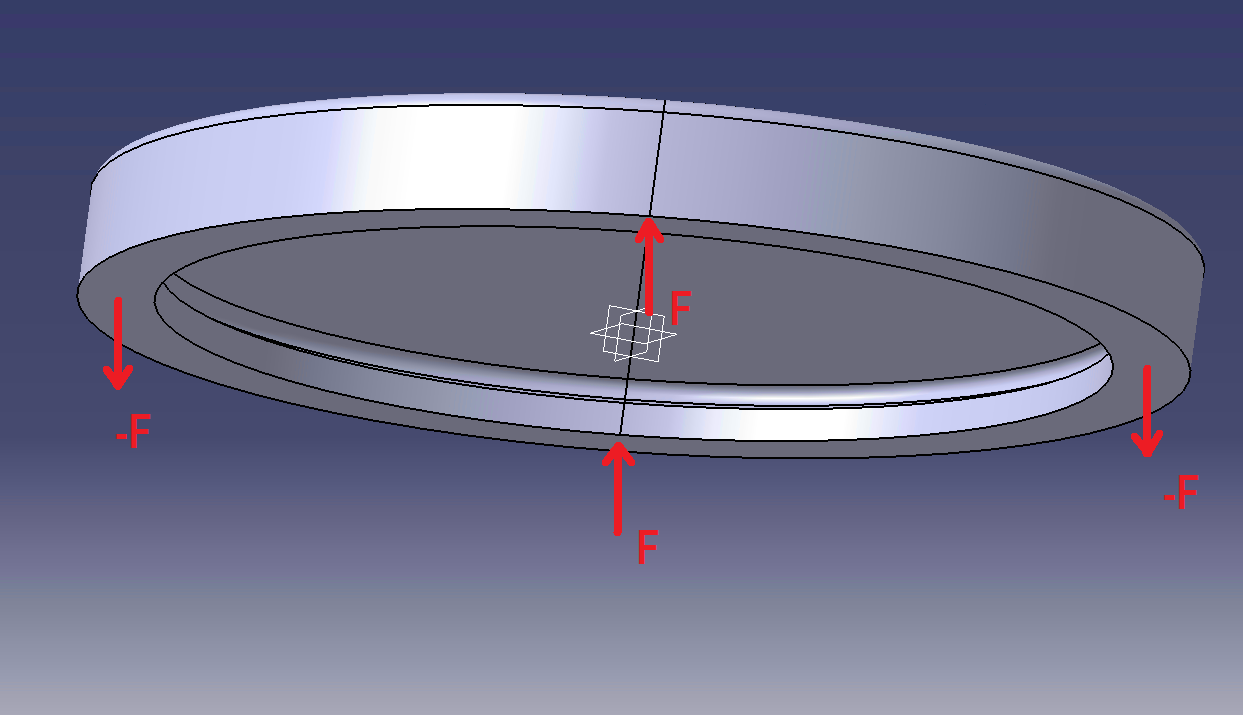}\hfill
  \includegraphics[height = 5cm]{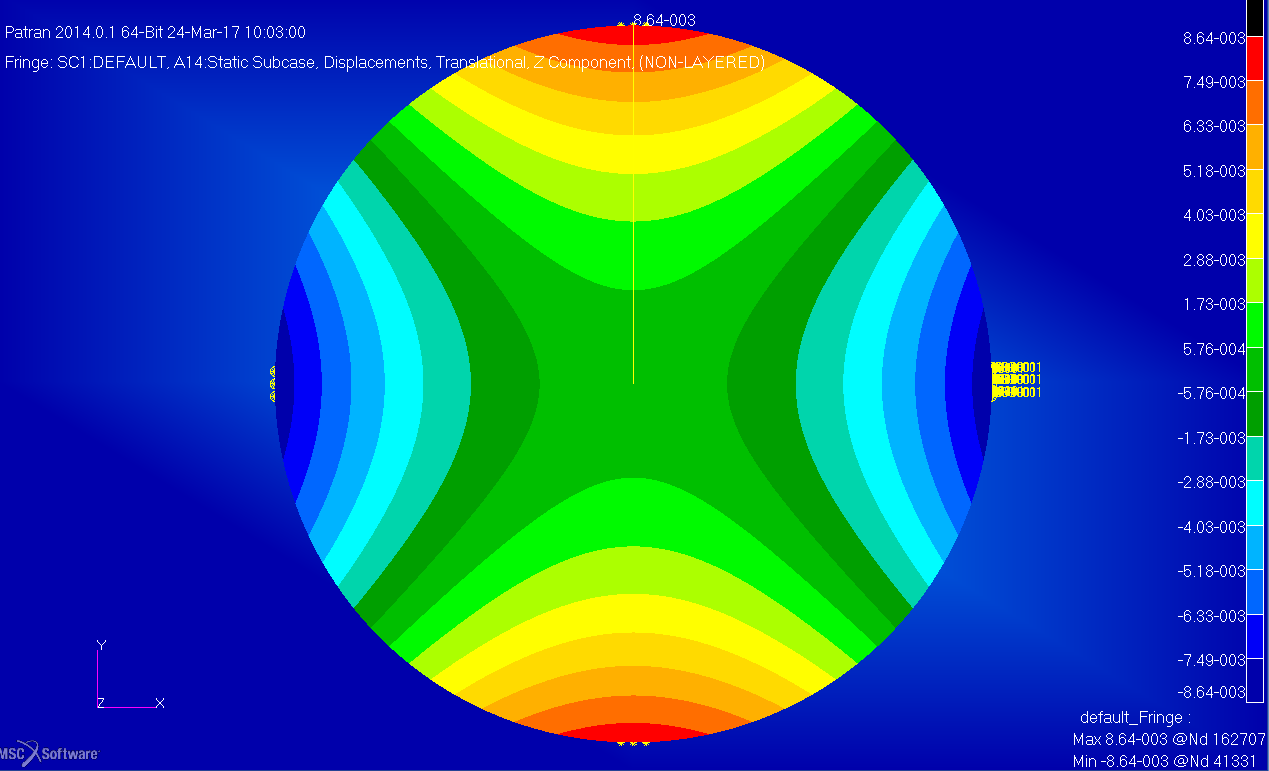}
  \caption{Left: CAD design of the substrate and the warping harness with the force application points. Right: Results of the nodes displacement after the FEA simulation.}
  \label{astig}
\end{figure}

After the simulation with Finite Element Analysis (FEA) we obtained the nodes displacement, Figure \ref{astig} (right).
The shape of astigmatism aberration is clearly shown. If we go further in computation by decomposing the surface shape with Zernike aberration (Table \ref{coeff}) in Matlab we obtain pure astigmatism with no residuals and a deformation of several microns, 3.365$\mu$ m with force application of 420N \cite{hugot:tel-00519452}.

\subsection{Non symmetrical harness}

\begin{figure}[H]
  \centering
  \includegraphics[height =5cm]{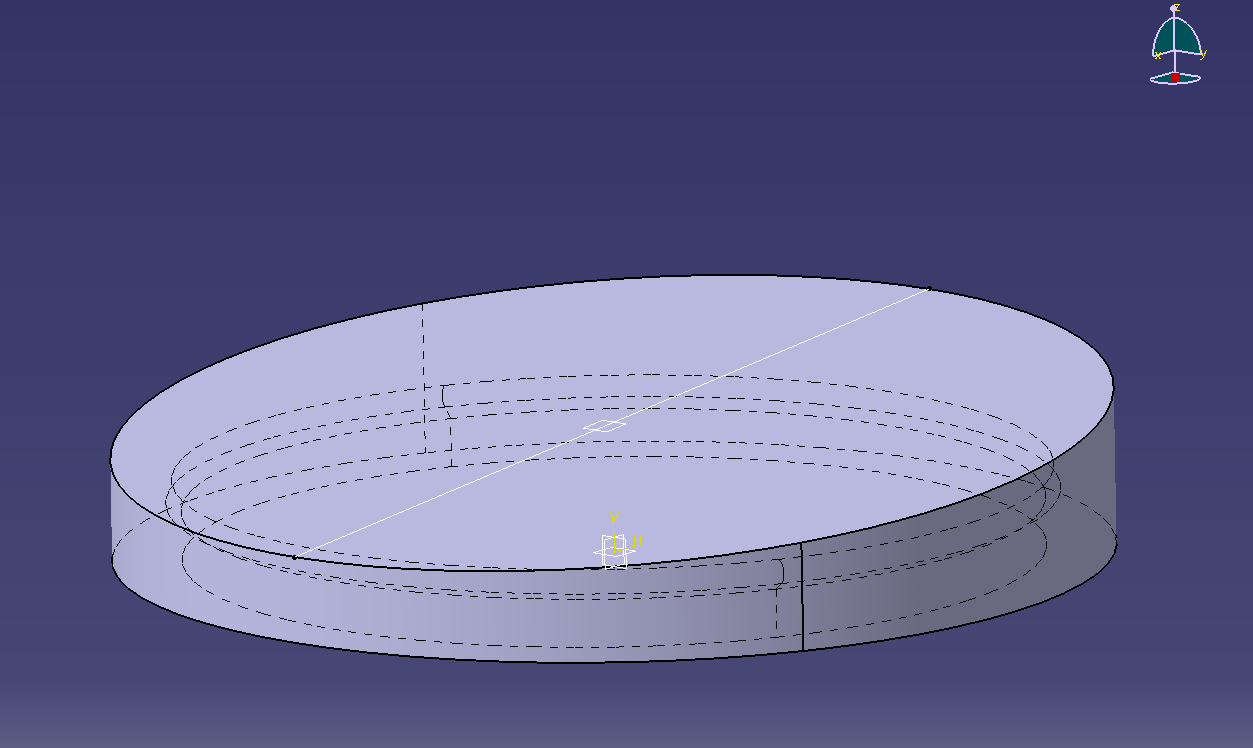}\hfill
  \includegraphics[height = 5cm]{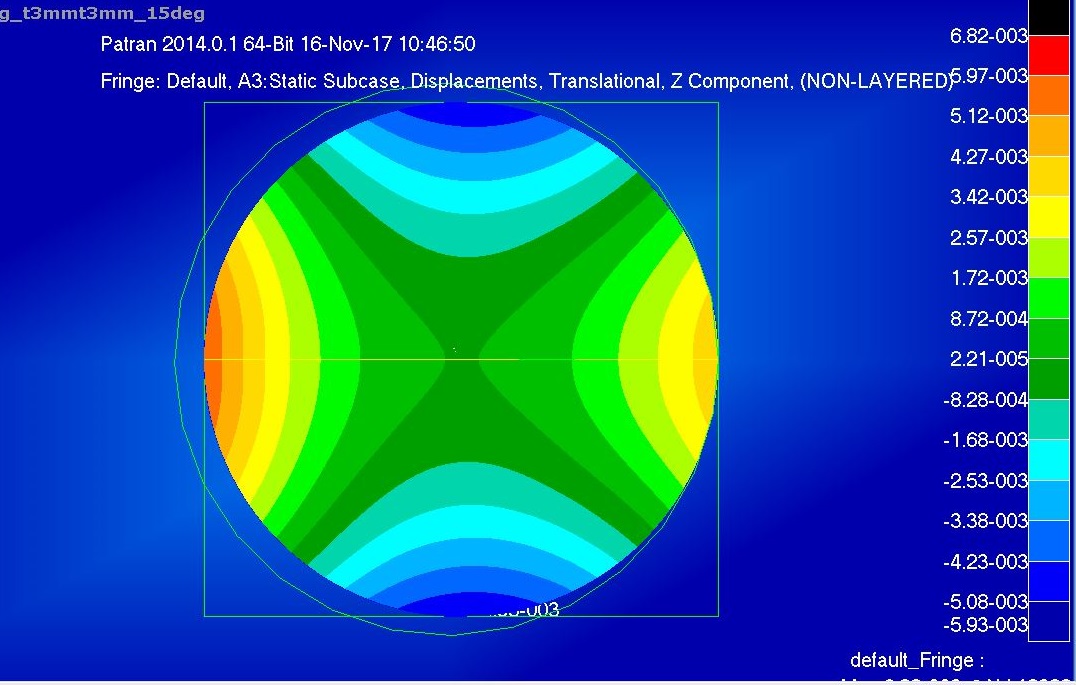}
  \caption{Left: CAD design of the substrate and the warping harness with the additional wedge. Right: Results of the nodes displacement after the FEA.}
  \label{wedge}
\end{figure}

In order to obtain the off axis parabola form, we need to break the symmetry of the warping harness to introduce coma aberration into the final deformation. To generate astigmatism and coma a wedge was added in the warping harness design, as shown in Figure \ref{wedge} (left).

After FEA the astigmatism pattern is still present, in Figure \ref{wedge} (right). The effect of the wedge is shown on the horizontal axis where the deformation is no longer symmetrical. 
Zernike decomposition has been performed and the amplitude of each Zernike terms is presented in Table \ref{NonsymmetricalResult}. 

\begin{table}[H]
\begin{center}
\begin{tabular}{|c|c|c|c|c|c|c|c|c|}
  \hline
  Aberrations & Piston & Tilt $y$ & Astig 3$x$ & Coma 3$y$ & Trefoil 5$y$ & Astig 5$x$ & Squa 7$x$ & Pent 7$x$\\
  \hline
  Amplitude $[\mu m]$ & $-0.064$ & $-0.354$ & $-2.847$ & $-0.109$ & $-0.112$ & $+0.029$ & $+0.009$ &$+0.011$\\
  \hline
  \end{tabular}
\caption{Results of the Zernike decomposition after FEA on the non symmetrical design}
  \label{NonsymmetricalResult}
  \end{center}
\end{table}

As expected we obtained coma 3, but if we compare to astigmatism 3 the ratio is 3.82\%, which does not yet meet the requirements and in addition, other non-desirable residuals are present. The magnitude of Trefoil 5 was of particular concern, as it shows high amplitude close to the amount of coma 3. Trefoil 5 aberration is undesirable in the surface shape because it will create errors in the final Point Spread Function (PSF) located in the disk where the coronagraph will detect exoplanets.

\subsection{Innovative warping harness}

\begin{figure}[H]
  \centering
  \includegraphics[height =6cm]{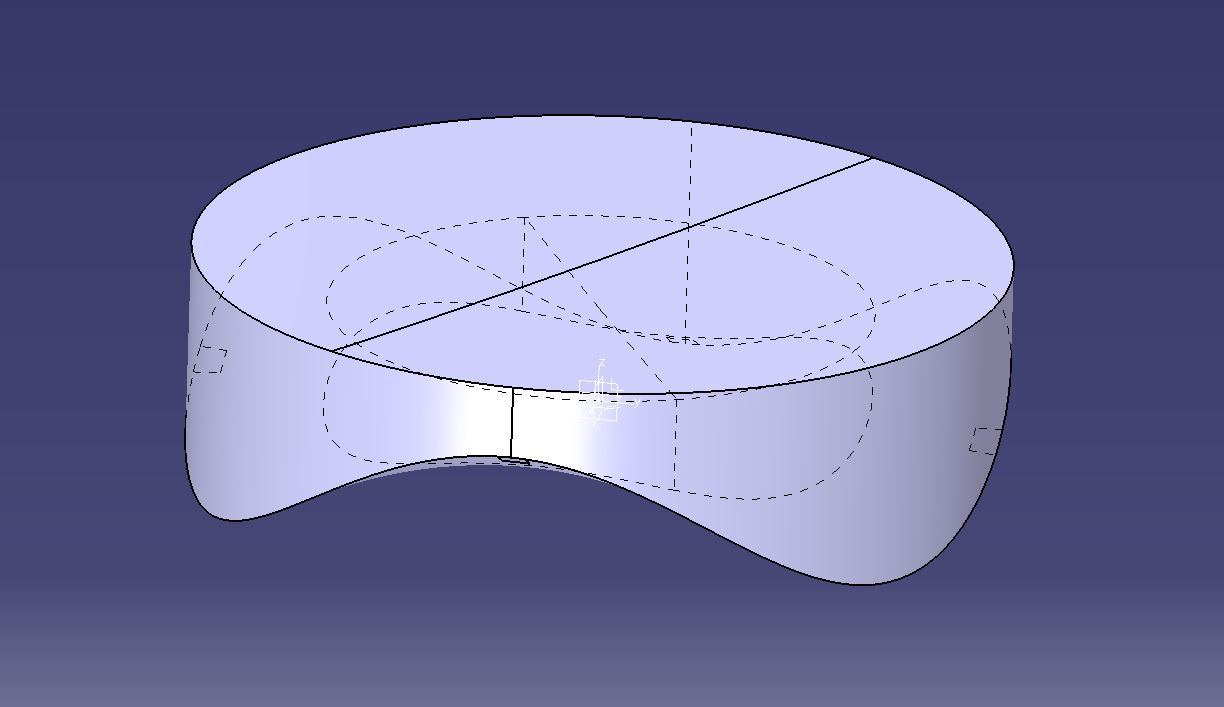}\hfill
  \includegraphics[height = 6cm]{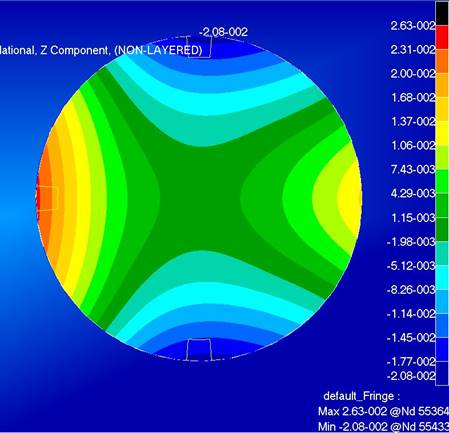}
  \caption{Left: CAD design of the substrate and the innovative warping harness. Right: Results of the nodes displacement after the FEA.}
  \label{innovative}
\end{figure}

Following the parametric study and topology optimisation \cite{bendsoe2004} we converged on an innovative warping harness with a thickness distribution optimised for the required off axis parabolas, in Figure \ref{innovative} (left). 

In Figure \ref{innovative} (right) is shown the node displacement after the deformation with FEA. As before there is still the astigmatism pattern, but with a larger non-symmetrical deformation on the horizontal axis which corresponds to coma 3 aberrations. With this design the ratio requirement is achieved, the $Ratio_{coma/astig}$ is 10.7\%, as shown by the ratio of coma 3 and astig 3 amplitudes in Table \ref{InnovativeResult}.

\begin{table}[H]
\begin{center}
\begin{tabular}{|c|c|c|c|c|c|c|c|c|}
  \hline
  Aberrations & Piston & Tilt $y$ & Astig 3$x$ & Coma 3$y$ & Trefoil 5$y$ & Astig 5$x$ & Squa 7$x$ & Pent 7$x$\\
  \hline
  Amplitude $[\mu m]$ & $-0.046$ & $-2.366$ & $-7.909$ & $-0.757$ & $-0.141$ & $+0.106$ & $+0.107$ & $+0.819$\\
  \hline
  \end{tabular}
\caption{Results of the Zernike decomposition after FEA on the innovative design}
  \label{InnovativeResult}
  \end{center}
\end{table}

\subsection{WFIRST second off axis parabola design}

This design has been applied to the second of the WFIRST off axis parabolas. The requirements for this off axis parabola in terms of deformation are $0.870\mu m$ in astigmatism 3 and $0.088\mu m$ in coma 3. Results of the deformation after FEA and Zernikes decomposition are shown in Table \ref{OAP2}. The deformation requirement are achieved, but the design generated additional residuals, which will be characterised through prototype production. This model will be our master piece for the set up of the fabrication process.

\begin{table}[H]
\begin{center}
\begin{tabular}{|c|c|c|c|c|c|c|c|c|}
  \hline
  Aberrations & Piston & Tilt $y$ & Astig 3$x$ & Coma 3$y$ & Trefoil 5$y$ & Astig 5$x$ & Squa 7$x$ & Pent 7$x$\\
  \hline
  Amplitude $[\mu m]$ & $-0.011$ & $-0.274$ & $-0.875$ & $-0.087$ & $-0.018$ & $+0.009$ & $+0.012$ &$+0.038$\\
  \hline
  \end{tabular}
\caption{Results of the Zernike decomposition after FEA on the second WFIRST off axis parabola}
  \label{OAP2}
  \end{center}
\end{table}

\section{PROTOTYPING}
\label{sec:proto} 

There are several options available in order to prototype the design.
\begin{figure}[H]
  \centering
  \includegraphics[height = 4cm]{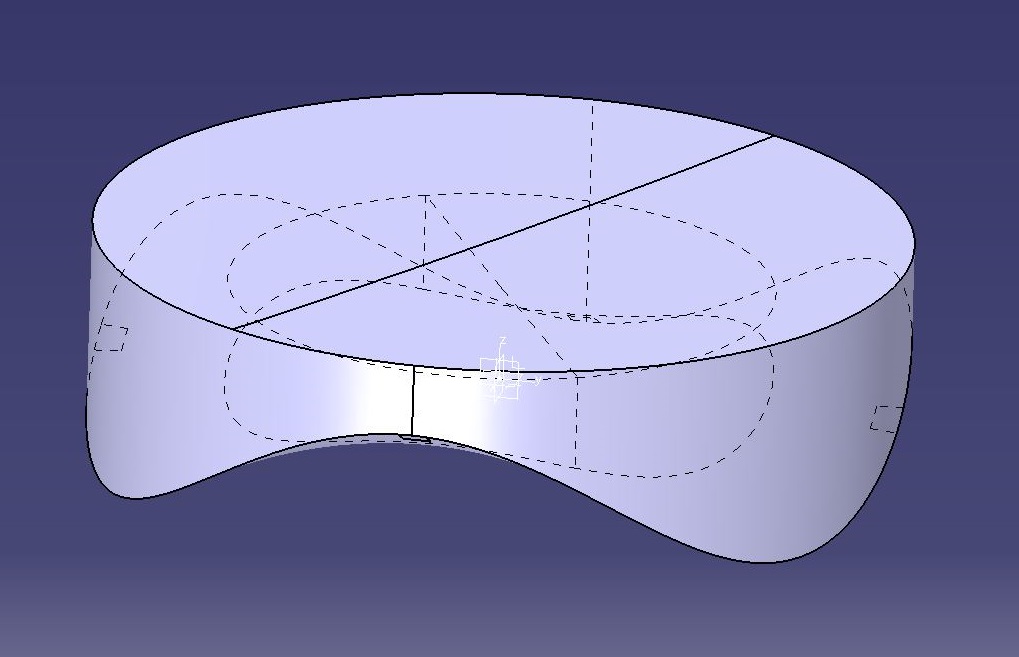}\hfill
  \includegraphics[height = 4cm]{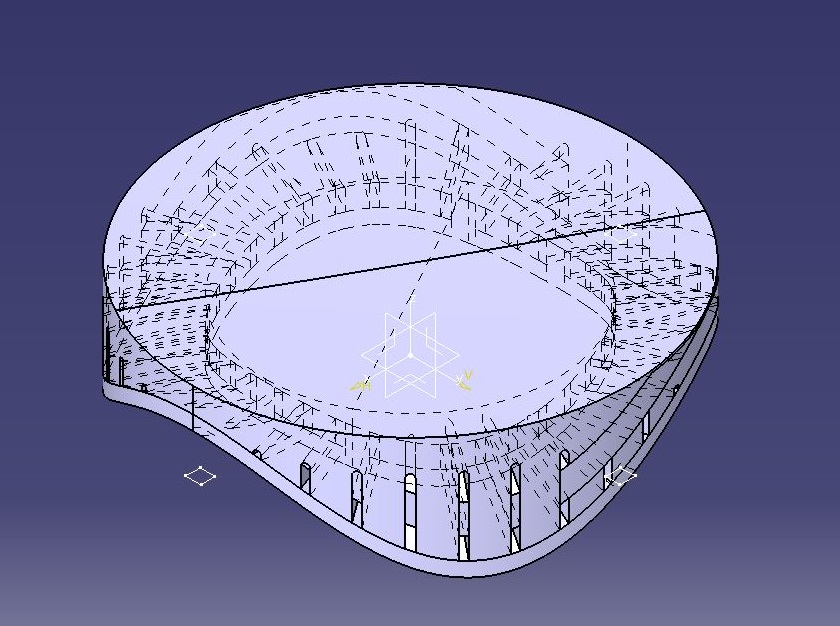}\hfill
  \includegraphics[height = 4cm]{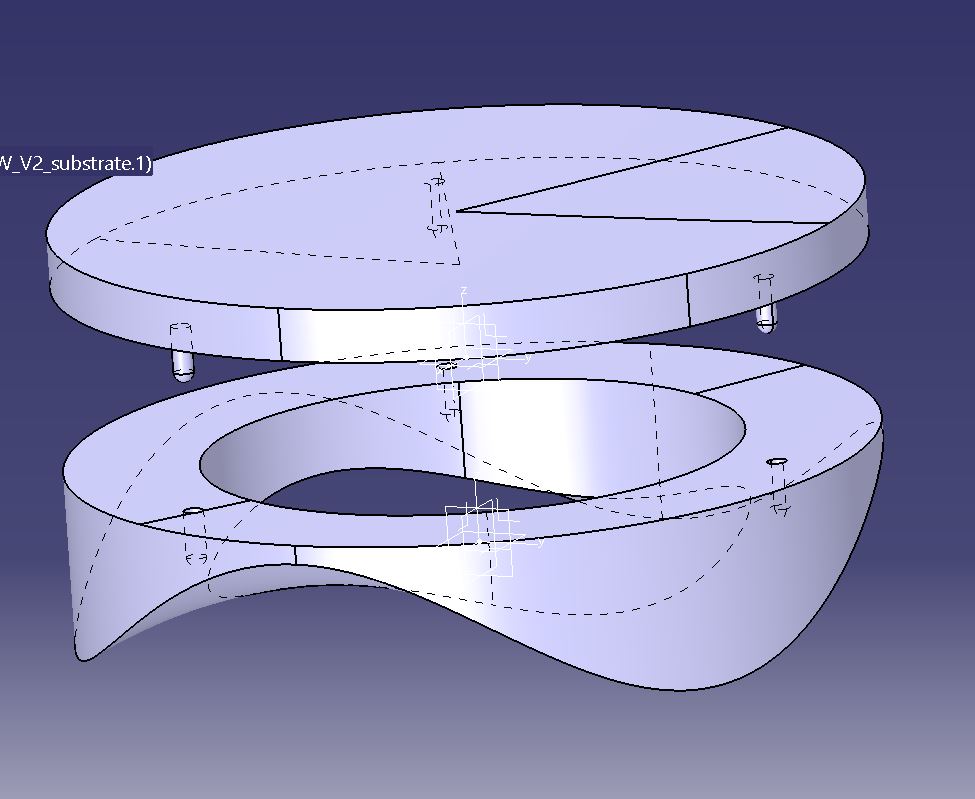}
  \caption{Left : CAD picture of the full model. Middle : CAD picture of the lightweight model. Right: CAD picture of the two-parts assembly model.}
  \label{prototype}
\end{figure}

The first option is that the off axis parabolas could be manufactured in one bloc of Zerodur, using traditional manufacturing, in Figure \ref{prototype} (left). This option is very expensive.

The second option will be studied in a future work to investigate lightweight structures and new materials, in Figure \ref{prototype} (middle). The off axis parabolas could be lightweight and printed by 3D printing. These prototypes aim to investigate the Corderite material and the possibility of polishing the surface. Corderite is a ceramic very close to Zerodur in terms of Coefficient of thermal expansion (CTE) and mechanical properties \cite{Schott263} \cite{Hirose1984}. During 3D printing the ceramic resin is photo-polymerised and then a cure is applied; this technique is called stereolithography \cite{Stampfl2014}. During this phase the maximal thickness should be $5mm$ to be sure that all the ceramic resin is cured, the lightweighting design offers a large surface area which aids cure, but in addition, it also minimises mass for launch weight restrictions. 
The most challenging point is this design is that the lightweight structure should impact as little as possible the deformation on the mirror substrate and still provide prefect off axis parabola.

The third option is the two-parts assembly model. Where the mirror substrate is manufactured in Zerodur by traditional manufacturing and the warping harness is printed. Multimaterial simulations have been done with several material options (Al, Ti, ceramic) for the harness and Zerodur for the substrate and there is no detrimental change on the surface deformation. So we can use different materials for the warping harness. The first prototype for WFIRST will be created via this method. The two-parts assembly model is the cheaper option.

\section{Conclusion}

In this study we created a warping harness design capable of generating off axis parabola deformation with only one actuator. The simulations have demonstrated the ability to manufacture off axis parabola using stress polishing combined with 3D printing. The development of this new manufacturing process shows that we can reach the WFIRST optical fabrication requirements with non-complex manufacturing.
Simulations demonstrate very good results in term of surface quality which is very encouraging for the prototyping phase. Residuals need to be characterised during this phase to remove them in the future harness designs or compensate for them with the support structure.
The initial prototypes are under fabrication and will soon be polished.

\acknowledgments 
 
The authors would like to acknowledge the European commission for funding this work through the Program H2020-ERC-STG-2015 – 678777 of the European Research Council.

\bibliography{report} 
\bibliographystyle{spiebib} 

\end{document}